\documentclass[12pt]{article}
\pdfoutput=1

\usepackage{amsmath}
\usepackage{mathtools}

\usepackage{booktabs,caption}
\usepackage[flushleft]{threeparttable}

\usepackage[square,sort,comma,numbers]{natbib}
\usepackage{setspace}
\usepackage[margin=1in]{geometry}

\usepackage{mathptmx}

\begin{document}

\title{An improved quality control chart to monitor the mean based on ranked sets}

\date{\today}
\author{}
\maketitle

\author{G.P. Silva\footnote{\label{note1}Department of Statistics, Federal University of Paran\'a, Curitiba, Brazil} \footnote{Corresponding author. Email: guilhermeparreira.silva@gmail.com}, C. A. Taconeli$^\text{\ref{note1}}$, W.M. Zeviani$^\text{\ref{note1}}$ and I. S. Guimarães$^\text{\ref{note1}}$}

\begin{abstract}
In this study, we considered the design and performance of control charts using neoteric ranked set sampling (NRSS) in monitoring normal distributed processes. NRSS is a recently proposed sampling design, based on the traditional ranked set sampling (RSS). We evaluated NRSS control charts by average run length (ARL), based on Monte Carlo simulation results. NRSS control charts performed the best, compared to RSS and some of its extensions, in most simulated scenarios. The impact of imperfect ranking was also evaluated. An application on strength concrete data serves as an illustration of the proposed method.

\end{abstract}

\vspace{1cm}


\section{Introduction}

Nowadays, technological resources are widely available for the real-time monitoring of many industrial processes. Even so, it must be recognized that sampling still plays a fundamental role in statistical quality control. Factors such as high costs or time of inspection and destructive tests may limit the evaluation of a large number of items. In this context, efficient sampling designs, providing more accurate results with smaller sample sizes, are highly useful. Sampling designs based in ranked sets have been shown as efficient alternatives to more conventional methodologies (such as simple random sampling) in industrial applications, particularly in developing statistical quality control charts.

Originally proposed in 1924 by Walter A. Shewart, statistical quality control charts (or simply control charts) constitute a relevant tool for visualizing industrial processes and identifying assignable causes of variation \cite{She24, Mon09}. A process is said to be under statistical control when no special or assignable causes are present. Several alternatives to the original control charts were proposed, providing greater speed in detecting out of control situations. These alternatives include: the use of additional or alternative decision rules \cite{WEC58, Kho03, Kou07}; adaptive sampling schemes \cite{Cos94, Tag98, Cos07}  or even the use of alternative sampling designs to the usual simple random sampling (SRS). In this study, we consider a variety of sampling designs based on ranked sets for constructing control charts.

Ranked set sampling (RSS) is an effective sampling design when the variable of interest is expensive or difficult to measure, but it is possible ranking sample units according to some accessible and cheap criterion \cite{Che03, Mci52}. This ranking process can be made based on an expert’s judgment or an auxiliary variable, strongly correlated to the variable of interest. The RSS design requires, initially, the selection of $k$ random samples (sets) of $k$ units. Then, the sample units are ranked within each set, according to the ordering criterion. From the $k^2$ original sample units, $k$ are effectively measured for the variable of interest (one from each set). RSS becomes more efficient than SRS as long as a more accurate and accessible ordering criterion is available. Several studies have shown the superiority of RSS over SRS for estimation of the  mean \cite{Del72, Tak68}, variance \cite{Mac02, Alh10}, among others population parameters (see also, for example, \cite{Alo14, Che07, Kau95}). Additionally, a large number of sampling designs derived from the original RSS were proposed, such as Median Ranked Set Sampling - MRSS \cite{Mut97}, Extreme Ranked Set Sampling - ERSS \cite{Sam96} and Double Ranked Set  Sampling - DRSS \cite{Als00}, among others \cite{Alo11a, Haq14a, Mut03a}.

Recently, RSS and its related sampling designs have been studied in the context of statistical quality control. \citep{Mut03b} developed different control charts using RSS and two of its modifications: ERSS and MRSS. Basically, ERSS differs from RSS by using the sample units ranked at the extremes (minimum and maximum), while MRSS is based only on the observations ranked in the central positions. The authors have shown, based on an extensive simulation study, that RSS-based control charts dominate their SRS counterpart, requiring, on average, fewer samples to detect a change in the process mean. Additionally, MRSS have showed the best performance among the three designs based on ranked sets.

Improved control charts for double-ranked set sampling schemes are presented in \cite{Abu04} and \cite{Alo12}. This class of sampling designs is characterized by the initial selection and ranking of $k^3$ (instead of $k^2$) sample units to draw a sample of size $k$ after two ranking cycles. Double ranked set sampling control charts outperform those based on a single ordering cycle. Additionally, the design and performance of cumulative sum and exponentially weighted moving average control charts in RSS can been seen, for example, in \cite{Als10, Haq14c, Haq15}. \cite{Alo14} present a bibliographic review of RSS based control charts.

Neoteric ranked set sampling (NRSS) \cite{Zam15} is a sampling design recently originated from RSS. Technically, its fundamental difference to RSS is the constitution and ordering of a single set of $k^2$ sample units. After the ordering process, $k$ units are chosen to compose the final sample, selected according to their ranks.  It was found, for different scenarios (sample sizes, correlation levels between the variable of interest and an auxiliary variable and probability distributions) that NRSS overcomes RSS and SRS for estimating population mean and variance. NRSS was firstly considered for control charts in \cite{Koy17} to monitor the mean of bivariate asymmetric distributions. The authors verified, through a Monte Carlo simulation study, that NRSS design has the best performance for skewed distributions when compared to others RSS based designs.

In this paper we propose and analyze control charts for monitoring the process mean based on NRSS for normal distributed processes. Its performance is evaluated by a Monte Carlo simulation study. We considered different sample sizes and processes with different shifts from statistical control. Also, we evaluated the impact of imperfect ranking by setting different correlation levels between the variable of interest and an auxiliary variable. NRSS control charts were compared with SRS, RSS, and some of its extensions based on results already presented in the literature. Finally, an example with concrete strength data complements this article.

\section{Ranked set sampling and other sampling designs based on ranked sets}

The RSS design can be described as follows:

\begin{enumerate}
\item Selection of $k^2$ units of the population using SRS, allocating them, randomly, in $k$ sets of size $k$;
\item Ranking the sample units in each set according to the possible values of the variable of interest, using the pre-established ordering criterion;
\item Selection, for the final sample, of the $i$th judged unit in the $i$th set, $i=1,2,...,k$.
\item Steps 1 to 3 can be replicated $n$ times ($n$ cycles) yielding a sample of size $nk$.
\end{enumerate}

We denote the sample by $Y_{[i]j}$ $(i=1,2,\dotsc,k; j=1,2,\dotsc,n)$, where $Y_{[i]j}$ represents the observation ranked in the $i$th position in the $j$th cycle. In this case, the sample units are independent, but not identically distributed random variables, as a result of the ordering process.

The usual estimator of the population mean using RSS is given by:
\begin{equation} 
\bar{Y}_{RSS}=\frac{1}{nk}\sum_{j=1}^{n}\sum_{i=1}^{k}Y_{[i]j},
\end{equation}

\noindent and its variance:
\begin{equation} 
Var(\bar{Y}_{RSS})=\frac{\sigma^{2}}{nk}-\frac{1}{nk^{2}}\sum_{j=1}^{n}\sum_{i=1}^{k}(\mu _{[i]}-\mu )^{2},
\end{equation}

\noindent where $\mu$ and $\sigma^2$ are the population mean and variance and $\mu_{[i]} = E  \left [Y_{[i]j}  \right ]$, that is the mean of the $i$th order statistic from a simple random sample of size $k$ under the perfect ranking scenario (when there are no errors in the ordering process).

As alternatives to the RSS we have, for example, the median ranked set sampling (MRSS), which consists in the selection of the judged median in each set. This sampling design provides an unbiased and more efficient estimator for the mean of symmetric distributions over RSS \cite{Mut97}.  Extreme ranked set sampling (ERSS), on the other hand, is based on the selection of the units judged as the minimum, in half of the sets, and the ones judged as the maximum, in the other half. This sampling design can be a convenient (although less efficient) alternative to RSS and MRSS, once it is simpler to identify extreme than intermediate ranked units \cite{Sam96}.

Furthermore, some sampling designs based on RSS starts from the initial draw of $k^3$ sample units in order to, after two ordering cycles, provide a final sample of $k$ units. Some examples of such designs are the quartile double ranked set sampling and the double quartile ranked set sampling; the extreme double ranked set sampling and the double extreme ranked set sampling; the median double ranked set sampling, the double median ranked set sampling and the double ranked set sampling  \citep{Als00}. Control charts based on these sampling designs perform better than those based on a single ordering cycle. However, they have higher costs as a counterpart, due to the selection and ordering of a higher number of sample units \citep{Abu04, Alo12}.

\section{Neoteric ranked set sampling}

Similar to the RSS, neoteric ranked set sampling (NRSS) is also useful when the ranking of sample units is much cheaper than obtaining their precise values \cite{Zam15}. NRSS design consists of the following steps:

\begin{enumerate}
\item Selection of $k^2$ units of the population using SRS;
\item Ranking the $k^2$ sample units based on the pre-established ordering criterion;
\item Selection of the $[(i-1)k+l]$-th sample unit for the final sample, $i=1, ..., \emph{k}$. If $k$ is odd, then $l=\frac{k+1}{2}$; if $k$ is even, then $l=\frac{k+2}{2}$ when $i$ is odd and $l=\frac{k}{2}$ when $i$ is even;
\item Again, steps 1-3 can be repeated $n$ times, setting up $n$ cycles and producing a final sample of size $nk$.
\end{enumerate}

In NRSS the $k^2$ original sample units compose (and are ordered in) a single set, which induces dependence between the observations (differently from the RSS design). The variances of these variables, however, are reduced due to the greater initial sample size, which justifies its higher efficiency. For illustration, to select a NRSS sample of size $k=3$, we would select the positions 2, 5 and 8 from a original sample of size $k^2 = 9$, ordered in a single set; for a sample of size $k=4$, we would select the positions 3, 6, 11 and 14 from a ordered sample of size $k^2 = 16$; and lastly, for a sample of size $k=5$ the positions 3, 8, 13, 18 and 23 would be selected from a ordered sample of size $k^2 = 25$. These are the sample sizes considered in this study. It is possible to observe that the positions of the selected sample units for the final sample are, in general terms, regularly spaced.

The NRSS sample is denoted by $\{Y_{[(i-1)k+l]j}; \hspace{0.1cm} i=1,2,...,k; \hspace{0.1cm} j=1,2,...n\}$, in which $Y_{[(i-1)k+l]j}$ refers to the unit ranked in position $[(i-1)k+l]$ (of an initial sample of size $k^2$), in the $j$th cycle. Under perfect ranking, particularly, $Y_{[(i-1)k+l]j}$ corresponds to the $((i-1)k+l)$th order statistics in a SRS of size $k^2$ taken from the population.

The NRSS sample mean is an unbiased estimator for the population mean for symmetric distributions, which can be written by:
\begin{equation}\label{eq:mean_NRSS}
\bar{Y}_{NRSS}=\frac{1}{nk}\sum_{j=1}^{n}\sum_{i=1}^{k}Y_{[(i-1)k+l]j},
\end{equation}

\noindent and its variance is given by:
\begin{equation}\label{eq:var_mean_NRSS}
Var(\bar{Y}_{NRSS})=\frac{1}{nk^{2}}\sum_{i=1}^{k}\text{Var}(Y_{[(i-1)k+l]})+\frac{2}{nk^{2}}\sum_{1\leq i<i'   \leq k}^{k}\text{Cov}(Y_{[(i-1)k+l]},Y_{[(i'-1)k+l]}).
\end{equation}

\section{Statistical quality control charts using NRSS (NRSS control charts)}

Control charts for the process mean based on simple random samples of size $k$ are defined by a central line (CL) and a pair of control limits (LCL and UCL) given by:
\begin{equation}\label{eq:LCL}
\begin{multlined}
LCL = \mu_{0} - A\sqrt{Var(\bar{Y}_{SRS})} = \mu_{0} -A\frac{\sigma_{0_{\bar{Y}_{SRS}}}}{\sqrt{k}}\\
\\
CL = \mu_{0}\\
\\
UCL = \mu_{0} + A\sqrt{Var(\bar{Y}_{SRS})} = \mu_{0} + A\frac{\sigma_{0_{\bar{Y}_{SRS}}}}{\sqrt{k}}
\end{multlined},
\end{equation}

\noindent where $\mu_{0}$ and $\sigma_{0}$ are the process mean and standard deviation under control state, $\bar{Y}_{SRS}$ the mean of a simple random sample of $k$ units, $A$ the amplitude parameter of the control chart and $\sigma_{0_{\bar{Y}_{SRS}}}$ the standard deviation of $\bar{Y}_{SRS}$. An observed sample mean beyond the control limits is an indicator of an out of control process.  It is usual to consider $A=3$, which, under normal distribution, is associated to a probability of a false alarm (a point outside the control limits for an under control process) of approximately 0.0027.

We propose control charts for the process mean using NRSS, based on the following limits:
\begin{equation}\label{eq:LCL_NRSS}
\begin{multlined}
LCL = \mu_{0} - A\sqrt{Var(\bar{Y}_{NRSS})} = \mu_{0} - A\sigma_{0_{\bar{Y}_{NRSS}}}\\
\\
CL = \mu_{0}    \\ 
\\
UCL = \mu_{0} + A\sqrt{Var(\bar{Y}_{NRSS})} = \mu_{0} + A\sigma_{0_{\bar{Y}_{NRSS}}}
\end{multlined},
\end{equation}

\noindent where $\mu_{0}$ is the mean of the process under control state, and $\bar{Y}_{NRSS}$ and $Var(\bar{Y}_{NRSS})$ are defined in (\ref{eq:mean_NRSS}) and (\ref{eq:var_mean_NRSS}), respectively.

Our proposal constitutes an extension of the conventional SRS control charts, in such a way that the samples are periodically selected using NRSS and the control limits are based on (\ref{eq:LCL_NRSS}). Alternatively, extensions of control charts were previously proposed for designs based on RSS. The performance of these control charts are used here as reference to NRSS control charts results.

In our study, to set the values for NRSS control limits, as described in (\ref{eq:LCL_NRSS}), it was firstly necessary to get the values for $Var(\bar{Y}_{NRSS})$ (for a process under statistical control) for each simulated scenario. Under perfect ranking, $Y_{[(i-1)k+l]}$ is equivalent to the $(i-1)k+l$ order statistic from a SRS sample of size $k^2$, $i=1,2,...,k$. So, in this case we calculated $Var(\bar{Y}_{NRSS})$ based on the variances and the covariances presents in (\ref{eq:var_mean_NRSS}). It was possible by using the distributions of order statistics, and its properties, presented, for example, in \cite{Bal98}.

Under imperfect ranking, on the other hand, due to the ranking errors, the sampling units no longer match to order statistics. In this case, we obtained the values for $Var(\bar{Y}_{NRSS})$ by means of a preliminary simulation study. So we simulated 1000000 NRSS samples from a bivariate normal distribution for different combinations of $k$ and $\rho$ (the correlation between the variable of interest and the auxiliary variable). Then, $Var(Y_{[(i-1)k+l]})$ and 
$Cov(Y_{[(i-1)k+l]}, Y_{[(i'-1)k+l]})$ were determined, respectively, by:

\begin{equation}\label{eq:varsimul}
Var(Y_{[(i-1)k+l]}) = \frac{\sum_{h=1}^{1000000} \left ( Y_{[(i-1)k+l],h} - \bar{Y}_{[(i-1)k+l]} \right )^2}
{1000000-1}, \hspace{0.3cm} i = 1,2,...,k,
\end{equation}

\noindent where 

\begin{equation}\label{eq:varsimulmean}
\bar{Y}_{[(i-1)k+l]} = \frac{\sum_{h=1}^{1000000} Y_{[(i-1)k+l],h}}{1000000}, 
\end{equation}

\noindent and

\begin{equation}\label{eq:covsimul}
\begin{split}
& Cov(Y_{[(i-1)k+l]}, Y_{[(i'-1)k+l]}) =\\ 
& =\frac{\sum_{h=1}^{1000000}\left ( Y_{[(i-1)k+l],h} - \bar{Y}_{[(i-1)k+l]} \right )
\left ( Y_{[(i'-1)k+l],h} - \bar{Y}_{[(i'-1)k+l]} \right )}
{1000000-1}, \hspace{0.3cm}
\end{split}
\end{equation}

\noindent for $1 \leq i < i' \leq k.$ Then, we replaced (\ref{eq:varsimul}) and (\ref{eq:covsimul}) in (\ref{eq:var_mean_NRSS}) to obtain the variances, and we used them to set the NRSS control limits under imperfect ranking. 

When the process parameters are unknown, we propose the estimation of $\mu_{0}$ and $Var(\bar{Y}_{NRSS})$ based on the results of $m$ samples of size $k$ selected from the process in the absence of assignable causes of variation (under control process), according to (\ref{eq:Est_Med_NRSS}) and (\ref{eq:Est_Var_NRSS}):
\begin{equation}\label{eq:Est_Med_NRSS}
\bar{\bar{Y}}_{NRSS}=\frac{1}{m} \sum_{p=1}^m \bar{Y}_{NRSSp}
\end{equation}
\noindent and
\begin{equation}\label{eq:Est_Var_NRSS}
\widehat{Var}\left ( \bar{Y}_{NRSS} \right ) = \frac{1}{k^2} \sum_{i=1}^k 
\widehat{Var} \left ( Y_{\left [ \left ( i-1 \right )k+l \right ]} \right )+
\frac{2}{k^2} \sum_{i < {i}'} 
\widehat{Cov} \left ( Y_{\left [ \left ( i-1 \right )k+l \right ]} , Y_{\left [ \left ( {i}'-1 \right )k+l \right ]}  \right ),
\end{equation}
\noindent so that 
\begin{equation}\label{eq:VarEstChap}
\widehat{Var} \left ( Y_{\left [ \left ( i-1 \right )k+l \right ]} \right ) = \frac{1}{m-1} \sum_{p=1}^m \left (Y_{\left [ \left ( i-1 \right )k+l \right ] p}- \bar{Y}_{\left [ \left ( i-1 \right )k+l \right ]} \right )^2, 
\end{equation}
\noindent where $\bar{Y}_{\left [ \left ( i-1 \right )k+l \right ]} = \frac{1}{m}\sum_{p=1}^m Y_{\left [ \left ( i-1 \right )k+l \right ]p} $ and
\begin{equation}
\begin{aligned}\label{eq:CovEstChap}
\widehat{Cov}(Y_{[(i-1)k+l]}, Y_{[( {i}'-1)k+l]}) = {}  & \frac{1}{m-1} \sum_{p=1}^m [(Y_{[( i-1)k+l]p}-\bar{Y}_{[( i-1 )k+l]})\\
                                                        & (Y_{[( {i}'-1)k+l]p} - \bar{Y}_{[( {i}'-1)k+l]})], 1\leq i < {i}' \leq k.
\end{aligned}
\end{equation}

So, in practice the NRSS control charts for the process mean with estimated control limits are defined by substituting, in (\ref{eq:LCL_NRSS}), $\mu_0$ by (\ref{eq:Est_Med_NRSS}) and $Var(\bar{Y}_{NRSS})$ by (\ref{eq:Est_Var_NRSS}):

\begin{equation}\label{eq:LCL_NRSS_Imp}
\begin{multlined}
LCL = \bar{\bar{Y}}_{NRSS} - A\sqrt{\widehat{Var}\left ( \bar{Y}_{NRSS} \right )} \\
\\
CL = \bar{\bar{Y}}_{NRSS}\\
\\
UCL = \bar{\bar{Y}}_{NRSS} + A\sqrt{\widehat{Var}\left ( \bar{Y}_{NRSS} \right )}
\end{multlined}.
\end{equation}

In order to investigate the bias of (\ref{eq:Est_Var_NRSS}) in estimating (\ref{eq:var_mean_NRSS}), an additional simulation study was carried out, considering $k=3,4 \text{ and } 5$. For each value of $k$, we simulated 50000 replications of $m$ samples, using NRSS, from a normal standard distribution. For $m$, values between 5 and 25 were considered. At each step, the $m$ simulated samples were considered to estimate (\ref{eq:Est_Var_NRSS}). We found that the bias of this estimator is negligible (a relative bias $\leq 0.001$ was verified for all considered sample sizes for $m$ $\geq$ 20).

\section{Simulation study}
\label{estudo_via_simulacao}

The performance of NRSS control charts was evaluated by a Monte Carlo simulation study. We simulated samples from a bivariate normal distribution, according to (\ref{eq:normal_biv}):
\begin{equation}\label{eq:normal_biv}
\binom{X}{Y}\sim \text{Normal}\left ( \binom{0}{\mu_{Y}} , \begin{pmatrix}
1 & \rho \\ 
\rho & 1
\end{pmatrix} \right ).
\end{equation}

We considered $\mu_Y=\mu_0=0$ as the under control process mean. For the out of control scenarios, we considered $\mu_Y=\mu_0+\delta \frac{\sigma_0}{\sqrt{k}}$, so that $\delta$ determines the shift in the process mean:
\begin{equation}\label{eq:delta}
\delta=\left | \mu_Y-\mu_0 \right |\frac{\sqrt{k}}{\sigma_0},
\end{equation}
\noindent such that $\delta = 0$ implies to an under control process.

\sloppy As parameters settings for the simulation study we considered $k=3, 4 \text{ and }5$;
$\delta = 0,0.1,0.2,0.3,0.4,0.8,1.2,1.6,2,2.4 \text{ and } 3.2$ and 
$\rho=0,0.25,0.50,0.75,0.9 \text{ and }1$. As the indicator for the performance of control charts we considered the Average Run Length (ARL), defined as the average number of points in a control chart until one exceeds the control limits. Particularly, if we have an under control process, $ARL_{0}$ is the reciprocal of the false alarm error rate; for an out of control process, $ARL_{1}$ is inversely proportional to the detection power, representing the average number of samples until the out of control state is detected.

For each combination of $k, \delta \text{ and } \rho$ we simulated 1000000 of NRSS samples, and the ARL values were calculated as the inverse of the proportion of points beyond the control limits.

The parameters for the simulation study were chosen in such a way to allow the comparison of the ARL values with those presented in other publications, referring to control charts for other sampling designs based on RSS. Moreover, it becomes evident that the considered scenarios (198 in total) comprises a great variety of processes. The sample size was limited in $k=5$ given the context for application of sampling designs based on RSS  (restrictions related to draw big samples, initial selection and ranking of $k^2$ - or even $k^3$ or more - sample units, among others).

Moreover, the amplitude parameter ($A$) of the control limits were set, under perfect ranking, so that $ARL_{0}=370.5$. This is the $ARL_{0}$ corresponding to SRS control charts when we set $A=3$. In this way, we could compare the $ARL_{1}$ values for NRSS control charts with those provided by other sampling designs. The double ranked set sampling designs control charts, particularly, produce low values for $ARL_{0}$ and, consequently, high false alarms rates when $A=3$. 

Tables from \ref{Tabp3}-\ref{Tabp5} present the simulated ARL for control charts with sample sizes $k=3, 4 \text{ and } 5$, respectively. Besides NRSS, results obtained by SRS, RSS, ERSS and MRSS are also presented. In this first part of the analysis, we considered perfect ranking $(\rho=1)$, allowing to assess the maximum gain possible in each design.


\begin{table}
\begin{threeparttable}
\caption{ARL for control charts constructed by designs based on RSS under perfect ranking when $k=3$.}
\centering
{\begin{tabular*}{\textwidth}{@{\extracolsep{\stretch{1}}}*{7}{r}@{}}
\toprule
$\delta$ & SRS & RSS\textsuperscript{a} & ERSS\textsuperscript{a} & MRSS & NRSS \\ 
  \midrule
0.00 & 370.40 & 370.51 & 370.51 & 370.51 & 370.51 \\ 
  0.10 & 352.93 & 333.89 & 340.25 & 339.67 & 325.63 \\ 
  0.20 & 308.43 & 266.03 & 272.18 & 265.11 & 234.03 \\ 
  0.30 & 253.14 & 196.93 & 197.20 & 186.22 & 157.23 \\ 
  0.40 & 200.08 & 139.43 & 137.99 & 128.12 & 102.60 \\ 
  0.80 & 71.55 & 35.43 & 35.35 & 29.52 & 21.25 \\ 
  1.20 & 27.82 & 11.54 & 11.43 & 9.22 & 6.41 \\ 
  1.60 & 12.38 & 4.76 & 4.75 & 3.80 & 2.76 \\ 
  2.00 & 6.30 & 2.50 & 2.49 & 2.06 & 1.61 \\ 
  2.40 & 3.65 & 1.61 & 1.61 & 1.40 & 1.20 \\ 
  3.20 & 1.73 & 1.09 & 1.09 & 1.04 & 1.01 \\ 
   \bottomrule
\end{tabular*}}
\label{Tabp3}
    \begin{tablenotes}
        \item{\textsuperscript{a}Source: Al-Omari and Haq \cite{Alo12}. \\ ARL values from SRS were calculated based on the sampling distribution of sample mean.}
\end{tablenotes}
\end{threeparttable}
\end{table}

\begin{table}[p]
\begin{threeparttable}
\caption{ARL for control charts constructed by designs based on RSS under perfect ranking when $k=4$.}
\centering
{\begin{tabular*}{\textwidth}{@{\extracolsep{\stretch{1}}}*{7}{r}@{}}
\toprule
$\delta$ & SRS & RSS\textsuperscript{a} & ERSS\textsuperscript{a} & MRSS & NRSS \\ 
  \midrule
0.00 & 370.40 & 370.51 & 370.51 & 370.51 & 370.51 \\ 
  0.10 & 352.93 & 328.08 & 341.30 & 318.07 & 310.56 \\ 
  0.20 & 308.43 & 249.81 & 266.81 & 232.45 & 210.30 \\ 
  0.30 & 253.14 & 174.89 & 192.64 & 156.42 & 126.90 \\ 
  0.40 & 200.08 & 119.36 & 135.85 & 100.29 & 77.86 \\ 
  0.80 & 71.55 & 27.78 & 33.69 & 21.42 & 13.89 \\ 
  1.20 & 27.82 & 8.54 & 10.70 & 6.38 & 4.09 \\ 
  1.60 & 12.38 & 3.55 & 4.41 & 2.73 & 1.89 \\ 
  2.00 & 6.30 & 1.94 & 2.33 & 1.59 & 1.25 \\ 
  2.40 & 3.65 & 1.35 & 1.53 & 1.19 & 1.06 \\ 
  3.20 & 1.73 & 1.03 & 1.07 & 1.01 & 1.00 \\ 
   \bottomrule
\end{tabular*}}
\label{Tabp4}
\begin{tablenotes}
        \item{\textsuperscript{a}Source: Al-Omari and Haq \cite{Alo12}. \\ ARL values from SRS were calculated based on the sampling distribution of sample mean.}
\end{tablenotes}
\end{threeparttable}
\end{table}

\begin{table}[p]
\begin{threeparttable}
\caption{ARL for control charts constructed by designs based on RSS under perfect ranking when $k=5$.}
\centering
{\begin{tabular*}{\textwidth}{@{\extracolsep{\stretch{1}}}*{7}{r}@{}}
\toprule
$\delta$ & SRS & RSS\textsuperscript{a} & ERSS\textsuperscript{a} & MRSS & NRSS \\ 
  \midrule
0.00 & 370.40 & 370.51 & 370.51 & 370.51 & 370.51 \\ 
  0.10 & 352.93 & 331.68 & 333.00 & 329.60 & 299.58 \\ 
  0.20 & 308.43 & 244.98 & 254.77 & 223.41 & 181.06 \\ 
  0.30 & 253.14 & 165.54 & 173.73 & 136.91 & 104.59 \\ 
  0.40 & 200.08 & 107.88 & 117.44 & 85.20 & 60.14 \\ 
  0.80 & 71.55 & 22.53 & 26.59 & 15.56 & 9.55 \\ 
  1.20 & 27.82 & 6.73 & 8.13 & 4.55 & 2.86 \\ 
  1.60 & 12.38 & 2.83 & 3.38 & 2.04 & 1.46 \\ 
  2.00 & 6.30 & 1.63 & 1.87 & 1.31 & 1.10 \\ 
  2.40 & 3.65 & 1.21 & 1.32 & 1.08 & 1.01 \\ 
  3.20 & 1.73 & 1.01 & 1.03 & 1.00 & 1.00 \\ 
   \bottomrule
\end{tabular*}}
\label{Tabp5}
\begin{tablenotes}
        \item{\textsuperscript{a}Source: Al-Omari and Haq \cite{Alo12}. \\ ARL values from SRS were calculated based on the sampling distribution of sample mean.}
\end{tablenotes}
\end{threeparttable}
\end{table}

Some conclusions drawn from tables \ref{Tabp3}-\ref{Tabp5} are the ones that follow:

\begin{itemize}

\item The efficiency of NRSS control charts for detecting shifts in process mean increases, as expected, for higher values of $\delta$ and $k$. As an illustration, for $k = 5$ we have $ARL = 60.14$ for $\delta = 0.40$ compared to $ARL = 2.86$ for $\delta = 1.20$, while for $\delta = 0.80$ we have $ARL = 21.25$ for $k = 3$ against  $ARL = 9.55$ for $k = 5$;

\item The NRSS control charts perform better than SRS control charts in all the simulated scenarios. For example, for $k = 3 \text{ and } \delta = 0.80$ we have $ARL = 21.25$ for NRSS control charts compared to $ARL = 71.55$ for SRS, while for $k = 5 \text{ and } \delta = 1.60$ we have $ARL = 1.46$ for NRSS against $ARL = 12.38$ for SRS;

\item The NRSS control charts dominates RSS and ERSS designs in all the simulated scenarios. For example, when compared to RSS, for $k = 3 \text{ and } \delta = 0.80$ we have $ARL = 21.25$ for NRSS control charts against $ARL = 35.43$ for RSS, while for $k = 5 \text{ and } \delta = 1.60$ we have $ARL = 1.46$ for NRSS against $ARL = 2.83$ for RSS;

\item The NRSS control charts overcome the MRSS competitor in all simulated scenarios. This is remarkable, once MRSS is well known by its higher efficiency in the mean estimation compared to RSS for symmetric distributions. Additionally,  MRSS performs best under both single and double ranked set strategies for control charts in some situations \cite{Meh13}. For $k = 3 \text{ and } \delta = 0.80$ it was verified $ARL = 21.25$ for NRSS control charts compared to $ARL = 29.52$ for MRSS, while for $k = 5 \text{ and } \delta = 1.60$ we have $ARL = 1.46$ for NRSS against $ARL = 2.04$ for MRSS.

\end{itemize}    

\begin{figure}[p]
\centering
\includegraphics[width=0.7\textwidth,angle=270]{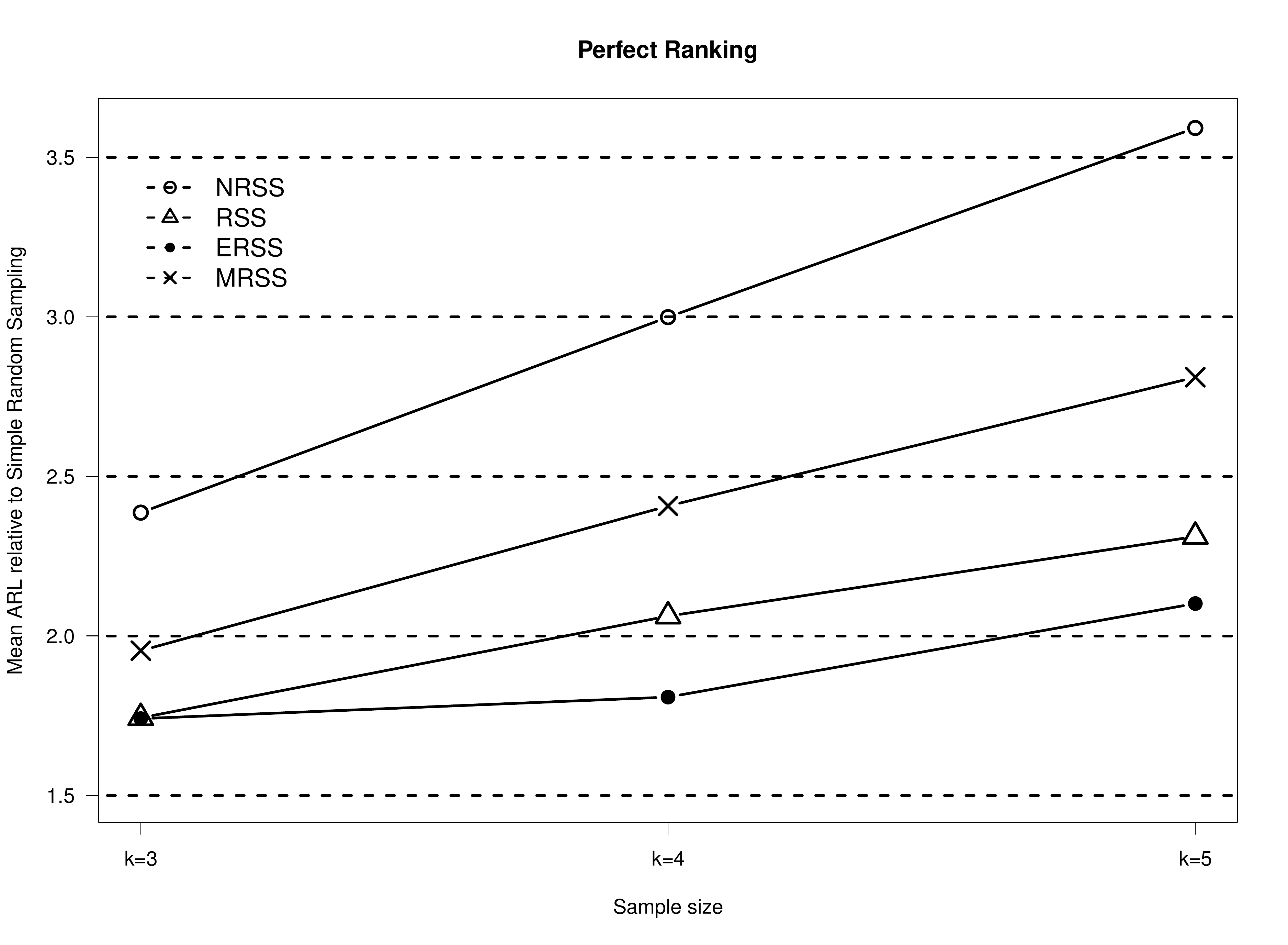}
\caption{Average relative efficiency from control charts of designs based on RSS compared to SRS under perfect ranking. ARL from RSS, MRSS and ERSS were taken from Al-Omari and Haq \cite{Alo12}.} \label{fig:fig1}
\end{figure}

In order to summarize the performance of the different control charts designs, Figure \ref{fig:fig1} presents the geometric means of the ratios of ARL values for SRS control charts  relative to the ones obtained by each of the other sampling designs, for each sample size. The ARL values for SRS control charts were, on average, 2.39 times larger than the corresponding NRSS when $k=3$; 3 times for $k=4$ and 3.59 times for $k=5$. The best performance of NRSS control charts over the RSS, ERSS and MRSS counterparts becomes evident. For MRSS, for example, we have, on average, ARL 1.22 times higher than NRSS for $k = 3$; 1.25 times for $k=4$ and 1.28 times for $k=5$. 


Tables \ref{table:k3} to \ref{table:k5} present the simulation results under imperfect ranking. The ARL values for $\rho=0$ are identical to the corresponding ones from SRS, once NRSS and SRS are equivalent if the ordering is completely random. Based on these results, it is possible to assess the impact of ranking errors in the performance of control charts. Additionally, for these simulations we decided to use the 3-sigma limits, by fixing $A=3$. We have two main reasons for this choice. First, this is a  traditional choice for Shewhart control charts, so it is possible to evaluate whether it is reasonable to set $A=3$ (particularly regarding the corresponding $ARL_{0}$ values). Furthermore, we verified in our simulation study that there is not a single $A$ value that reaches a specific $ARL_0$ for every situation. So, it could be difficult, for practical purposes, set up different values for this constant according to each industrial process.

    
\begin{table}[ht]
\caption{ARL for NRSS control charts under imperfect ranking when $k=3$.}
\centering
{\begin{tabular*}{0.9\textwidth}{@{\extracolsep{\fill} } rrrrrrrr}
  \hline
    &  \multicolumn{6}{c}{$\rho$}\\
  \cline{2-7}
\multicolumn{1}{c}{$\delta$} & \multicolumn{1}{c}{0.00} & \multicolumn{1}{c}{0.25} & \multicolumn{1}{c}{0.50} & \multicolumn{1}{c}{0.75} & \multicolumn{1}{c}{0.90} & \multicolumn{1}{c}{1.00} \\ 
  \hline
0.00 & 370.40 & 375.66 & 389.86 & 363.64 & 362.32 & 378.21 \\ 
 0.10 & 352.93 & 346.02 & 355.11 & 329.82 & 326.37 & 333.00 \\ 
 0.20 & 308.43 & 306.47 & 303.03 & 278.01 & 258.73 & 232.99 \\ 
 0.30 & 253.14 & 245.46 & 241.49 & 211.69 & 179.86 & 154.01 \\ 
 0.40 & 200.08 & 191.35 & 184.60 & 157.48 & 126.44 & 103.10 \\ 
 0.80 & 71.55 & 68.11 & 59.69 & 43.59 & 31.15 & 21.61 \\ 
  1.20 & 27.82 & 26.27 & 22.26 & 15.10 & 9.93 & 6.46 \\ 
 1.60 & 12.38 & 11.58 & 9.60 & 6.37 & 4.16 & 2.76 \\ 
 2.00 & 6.30 & 5.91 & 4.87 & 3.26 & 2.23 & 1.61 \\ 
  2.40 & 3.65 & 3.43 & 2.86 & 2.01 & 1.49 & 1.20 \\ 
 3.20 & 1.73 & 1.65 & 1.46 & 1.19 & 1.06 & 1.01 \\ 
   \hline
\end{tabular*}}
\label{table:k3}
\end{table}

\begin{table}[p]
\caption{ARL for NRSS control charts under imperfect ranking when $k=4$.}
\centering
{\begin{tabular*}{0.8\textwidth}{@{\extracolsep{\fill} } rrrrrrrr}
  \hline
    &  \multicolumn{6}{c}{$\rho$}\\
  \cline{2-7}
\multicolumn{1}{c}{$\delta$} & \multicolumn{1}{c}{0.00} & \multicolumn{1}{c}{0.25} & \multicolumn{1}{c}{0.50} & \multicolumn{1}{c}{0.75} & \multicolumn{1}{c}{0.90} & \multicolumn{1}{c}{1.00} \\ 
  \hline
0.00 & 370.40 & 353.48 & 362.06 & 375.94 & 372.44 & 378.50 \\ 
  0.10 & 352.93 & 365.36 & 348.43 & 342.47 & 333.56 & 314.76 \\ 
  0.20 & 308.43 & 313.38 & 293.51 & 281.93 & 245.64 & 204.67 \\ 
  0.30 & 253.14 & 250.00 & 229.10 & 203.13 & 170.62 & 129.02 \\ 
  0.40 & 200.08 & 193.65 & 177.78 & 148.28 & 110.07 & 77.56 \\ 
  0.80 & 71.55 & 67.66 & 57.44 & 40.60 & 24.86 & 13.94 \\ 
  1.20 & 27.82 & 26.16 & 21.31 & 13.29 & 7.67 & 4.09 \\ 
  1.60 & 12.38 & 11.60 & 9.14 & 5.59 & 3.23 & 1.88 \\ 
  2.00 & 6.30 & 5.85 & 4.67 & 2.89 & 1.82 & 1.25 \\ 
  2.40 & 3.65 & 3.41 & 2.75 & 1.82 & 1.29 & 1.06 \\ 
  3.20 & 1.73 & 1.64 & 1.42 & 1.14 & 1.02 & 1.00 \\ 
   \hline
\end{tabular*}}
\label{table:k4}
\end{table}

\begin{table}[p]
\caption{ARL for NRSS control charts under imperfect ranking when $k=5$.}
\centering
{\begin{tabular*}{0.8\textwidth}{@{\extracolsep{\fill} } rrrrrrrr}
  \hline
    &  \multicolumn{6}{c}{$\rho$}\\
  \cline{2-7}
\multicolumn{1}{c}{$\delta$} & \multicolumn{1}{c}{0.00} & \multicolumn{1}{c}{0.25} & \multicolumn{1}{c}{0.50} & \multicolumn{1}{c}{0.75} & \multicolumn{1}{c}{0.90} & \multicolumn{1}{c}{1.00} \\ 
  \hline
0.00 & 370.40 & 359.20 & 371.89 & 372.44 & 379.51 & 379.65 \\ 
  0.10 & 352.93 & 341.41 & 346.50 & 345.18 & 324.78 & 299.85 \\ 
  0.20 & 308.43 & 303.03 & 293.00 & 278.09 & 236.07 & 181.88 \\ 
  0.30 & 253.14 & 246.43 & 230.63 & 197.63 & 155.13 & 102.83 \\ 
  0.40 & 200.08 & 192.98 & 181.39 & 142.51 & 101.27 & 59.59 \\ 
  0.80 & 71.55 & 66.90 & 57.01 & 37.25 & 21.06 & 9.61 \\ 
  1.20 & 27.82 & 25.89 & 20.65 & 12.27 & 6.35 & 2.87 \\ 
  1.60 & 12.38 & 11.45 & 8.95 & 5.11 & 2.72 & 1.46 \\ 
  2.00 & 6.30 & 5.78 & 4.54 & 2.67 & 1.59 & 1.10 \\ 
  2.40 & 3.65 & 3.38 & 2.68 & 1.71 & 1.19 & 1.01 \\ 
  3.20 & 1.73 & 1.63 & 1.40 & 1.11 & 1.01 & 1.00 \\ 
   \hline
\end{tabular*}}
\label{table:k5}
\end{table}

\begin{itemize}

\item Control charts for all designs based on RSS lose performance when the correlation between the variables decreases. For example, for NRSS control charts, $k=3$ and $\delta = 0.8$, $ARL=31.15$ when $\rho = 0.90$; $43.59$ when rho $\rho = 0.75$ and 59.69 when $\rho = 0.50$;

\item The ARL values for NRSS control charts are smaller compared to the ones produced by SRS in 147 of the 150 simulated scenarios with $\delta \neq 0$. NRSS only loses in three scenarios described by low shifts in process mean and low values for $\rho$;

\item The $ARL_{0}$ values from NRSS control charts oscillated around $370.5$, as intended. On average, we have $ARL_{0}=373.94$ for $k=3$; $ARL_{0}=368.48$ for $k = 4$ and $ARL_{0}=372.54$ for $k=5$, with individual values of $ARL_{0}$ varying from 353.48, when $k=4 \text{ and } \rho =0.25$, to 389.86, when $k=3 \text{ and } \rho =0.50$.

\end{itemize}

\begin{figure}[htb]
\centering
\includegraphics[width=0.7\textwidth,angle=270]{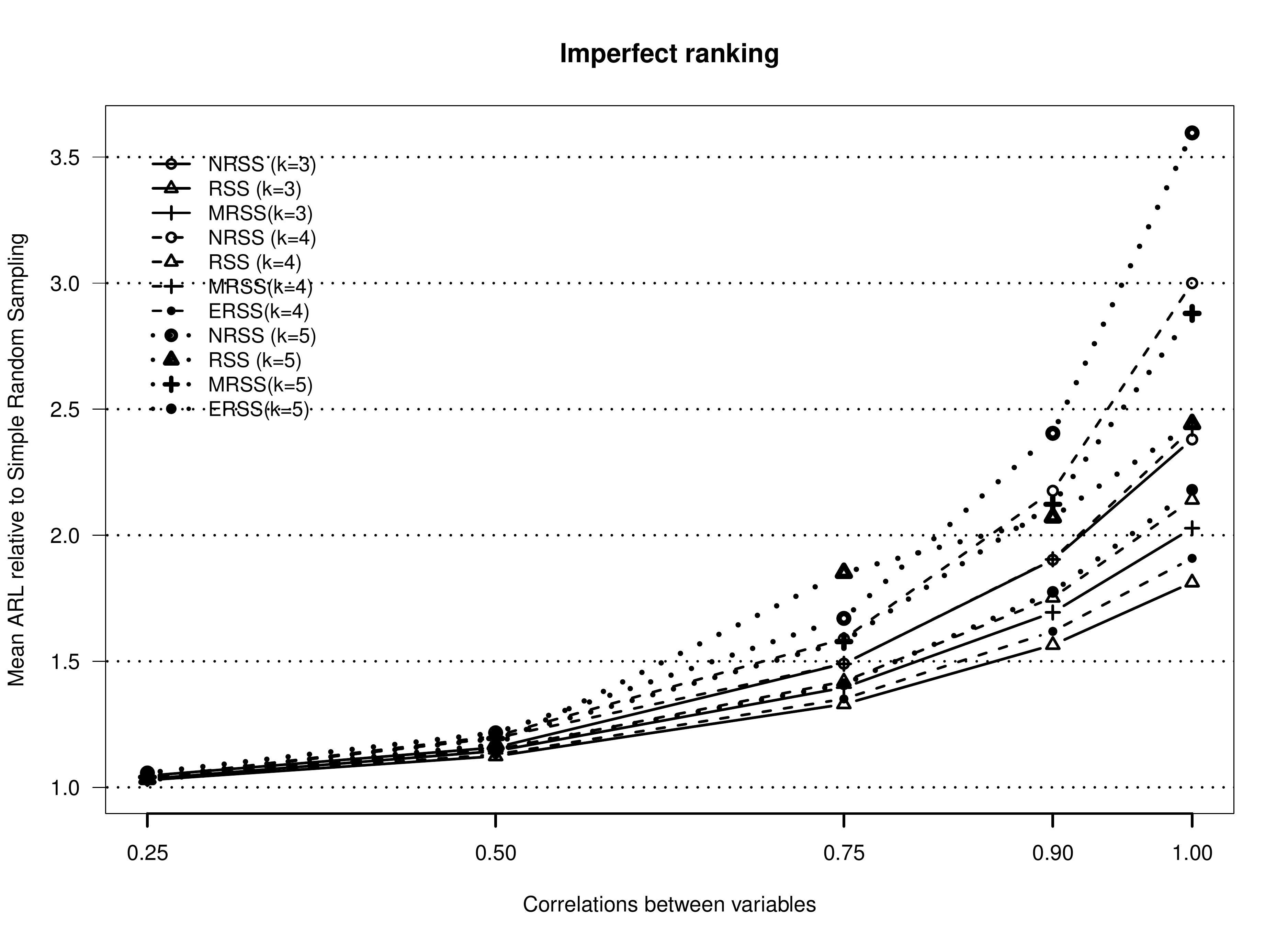}
\caption{Average relative efficiency from control charts of designs based on RSS compared to SRS under imperfect ranking. ARL from RSS, MRSS and ERSS were taken from Al-Omari and Haq \cite{Alo12}. For $k=3$ RSS and ERSS provides the same sampling design.}  \label{fig:fig2}
\end{figure}

Figure \ref{fig:fig2} shows the geometric means of the ratios of ARL values for SRS control charts relative to the ones obtained by each of the RSS based designs. These results are presented for each sample size and considering the different correlation levels between the auxiliary and the variable of interest. It is possible to observe that the NRSS control charts are, in general, more efficient than all other considered sampling designs. Moreover, the superiority of NRSS control charts becomes more accentuated when the correlation between the variables becomes higher. For $\rho = 0.9$ and $\rho = 1$, we have, on average, higher efficiency for the NRSS control charts with $k=4$ than for the other sampling designs taking $k=5$, which can reflect in resource savings and lower operational effort.


\section{Construction of control charts using real data}

In order to illustrate the application of the proposed methodology, we used a dataset with 1030 observations about the concrete strength to compression (MPa)  and the amount of cement (kg) used in the production of concrete blocks \cite{yeh98}. Although this data was not recorded as a case of a quality control process, it serves us, under some assumptions, as reference population, from which samples were drawn and control charts were constructed. We assumed the concrete strength as the variable of interest and the amount of cement as an auxiliary variable. Moreover, we assumed the concrete blocks strength distribution in this sample as the natural variability of an industrial process. 

In this application, we considered three sampling designs: SRS, RSS and NRSS; two sample sizes: $k = 3$ and $k = 5$, and processes in two different scenarios: under control ($\delta=0$) and out of control, considering $\delta=1.2$, as described in (\ref{eq:delta}). Under each sampling design and for each sample size, we selected, with replacement, 25 samples from the original data. These samples were considered for estimating the control limits (phase 1). Afterwards, 75 new samples were selected for monitoring the process mean (phase 2). For $\delta=0$, these 75 samples were selected with replacement from the original data; for $\delta=1.2$, we added to the strength values a normal random variable with mean $1.2  \frac{\sigma_{0}}{\sqrt{k}}$ and standard deviation equals to 2 (corresponding to 11.97\% of the concrete strength standard deviation). This standard deviation value is small enough to characterise the lack of control, predominantly, due to the shift in the process mean, in detriment to its dispersion (variance).

\begin{figure}[ht!]
\centering
\includegraphics[width=1\textwidth]{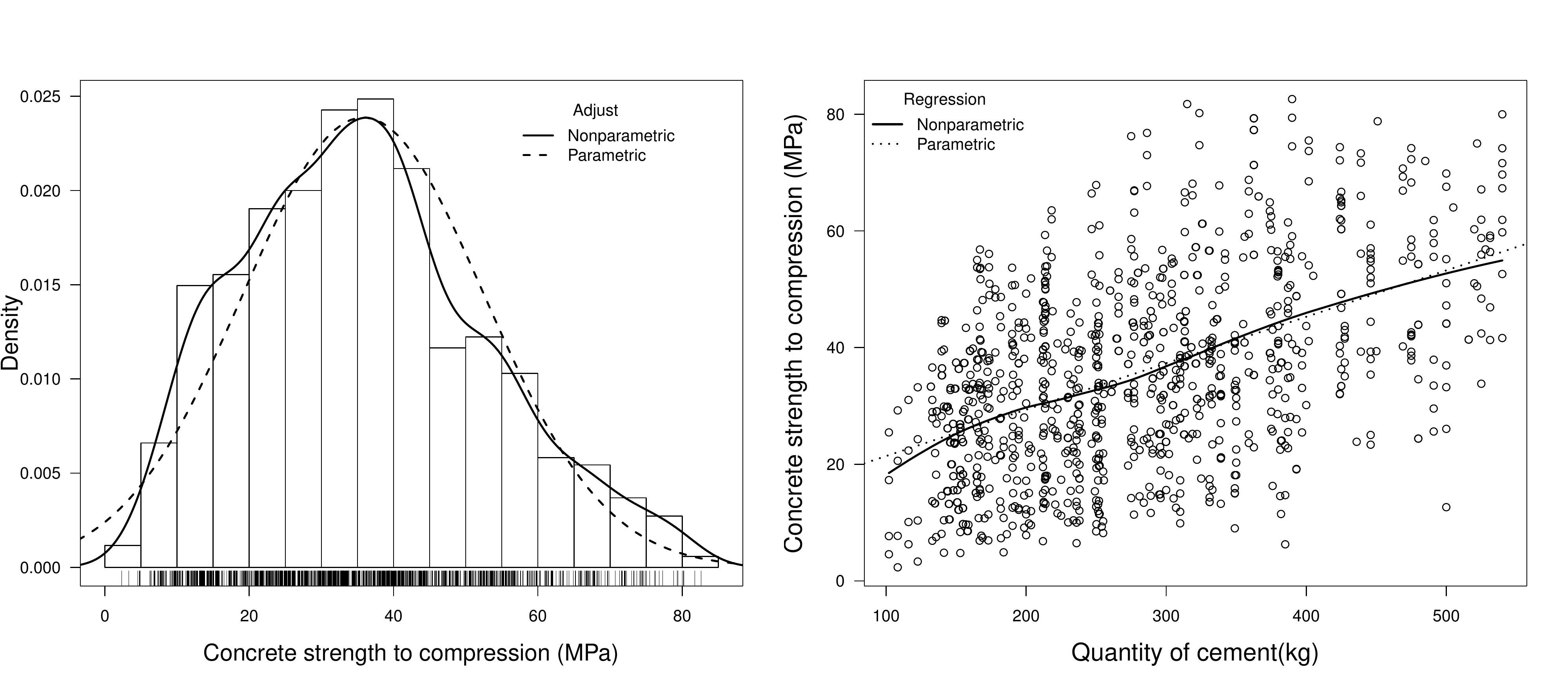}
\caption{Histogram and scatter plot for concrete strength.} \label{histcorr}
\end{figure}

Figure \ref{histcorr} presents (on the left) the histogram for the distribution of concrete strength, with the estimated normal distribution and Kernel density curves. The dispersion plot, on the right, indicates moderate positive linear relationship between the variables. The linear correlation coefficient is $\rho = 0.50$, which points to a moderately favourable scenario for designs based on RSS.


\begin{figure}[ht!]
\centering
\includegraphics[width=1\textwidth]{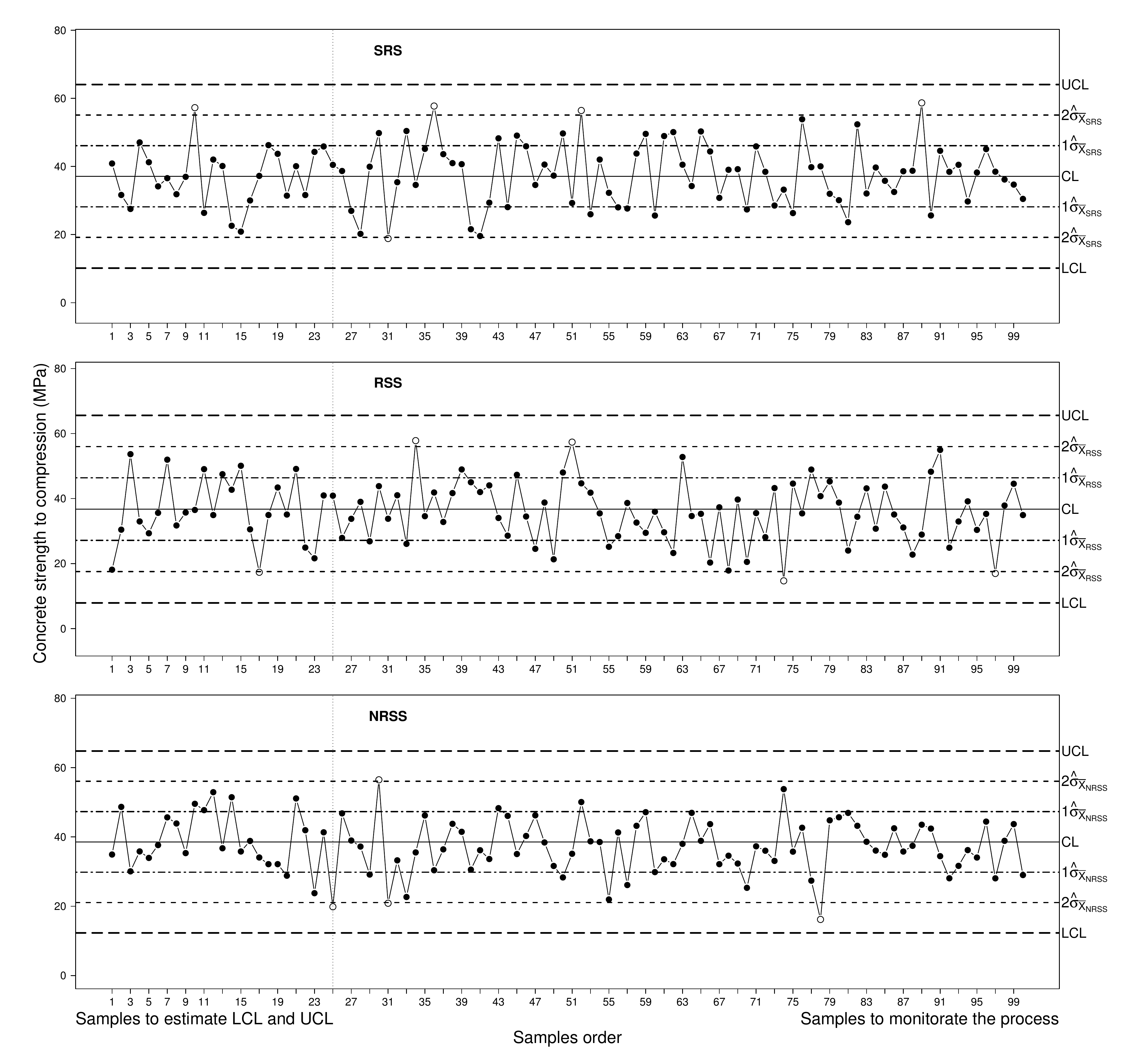}
\caption{Control charts for concrete strength considering $k=3$ and an under control process ($\delta = 0$)} \label{k3delta0}
\end{figure}

\begin{figure}[ht!]
\centering
\includegraphics[width=1\textwidth]{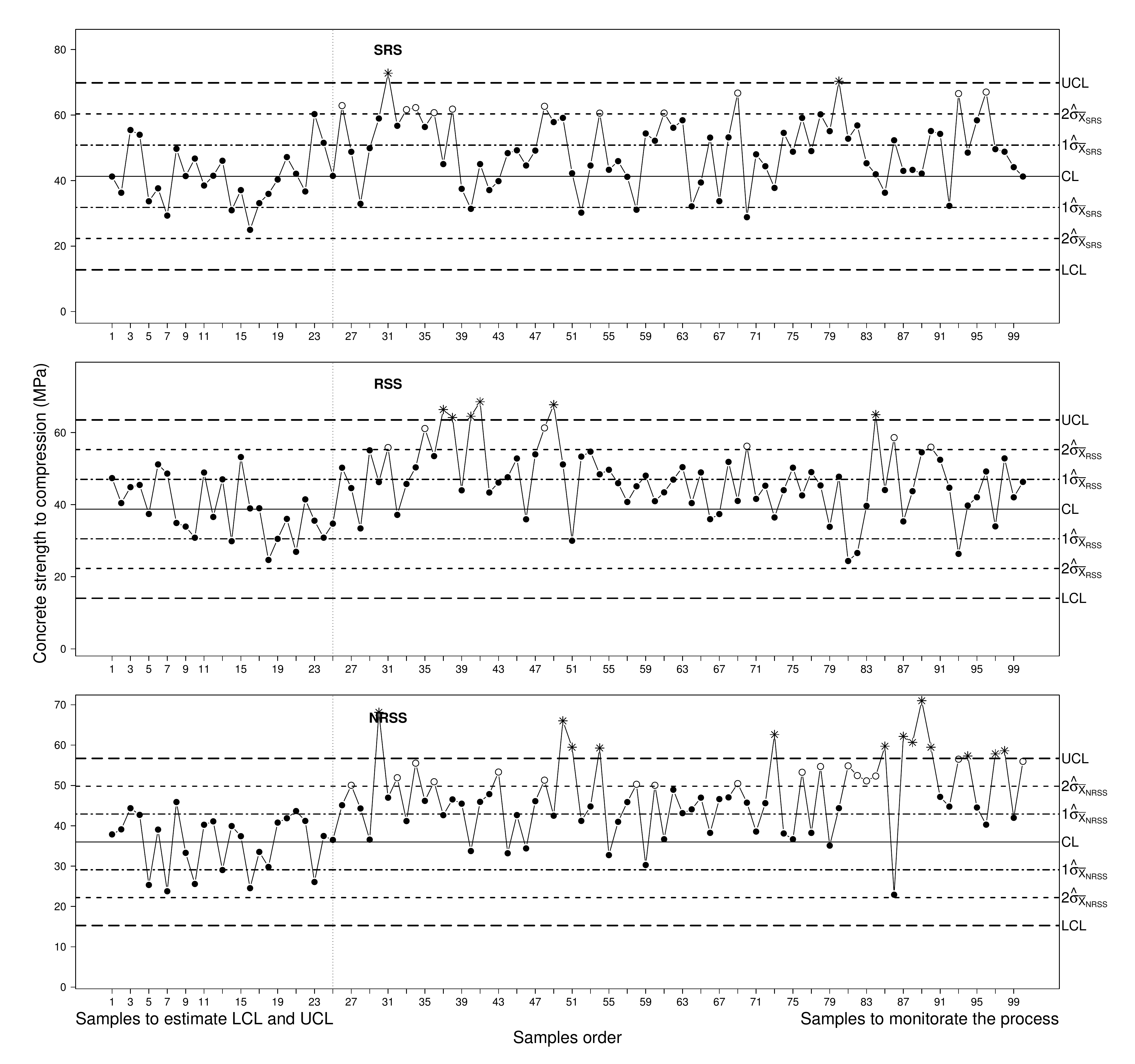}
\caption{Control charts for concrete strength considering $k=3$ and an out of control process ($\delta = 1.2$)} \label{k3delta12} 
\end{figure}

Following, Figures \ref{k3delta0} and \ref{k3delta12} present the SRS, RSS and NRSS control charts for the mean considering $k=3$. In Figure \ref{k3delta0} we have the charts produced considering $\delta = 0$ (under control process). For the three sampling designs, it is possible to observe points randomly distributed around the central line, without any point exceeding the control limits. This behaviour characterises an under control process, as expected. On the other hand, Figure \ref{k3delta12} presents the control charts for $\delta=1.2$ (out of control process). It is possible to observe that the NRSS control chart showed the highest number of points exceeding the control limits (13), followed by RSS (with 6 points outside the limits) and SRS control charts (only 2 points outside the limits). Moreover, we notice a higher number of points below the central line in the SRS control charts compared to its contenders. This indicates greater difficulty in lack of control detection for SRS than for RSS or NRSS control charts. 


\begin{figure}[ht!]
\centering
\includegraphics[width=1\textwidth]{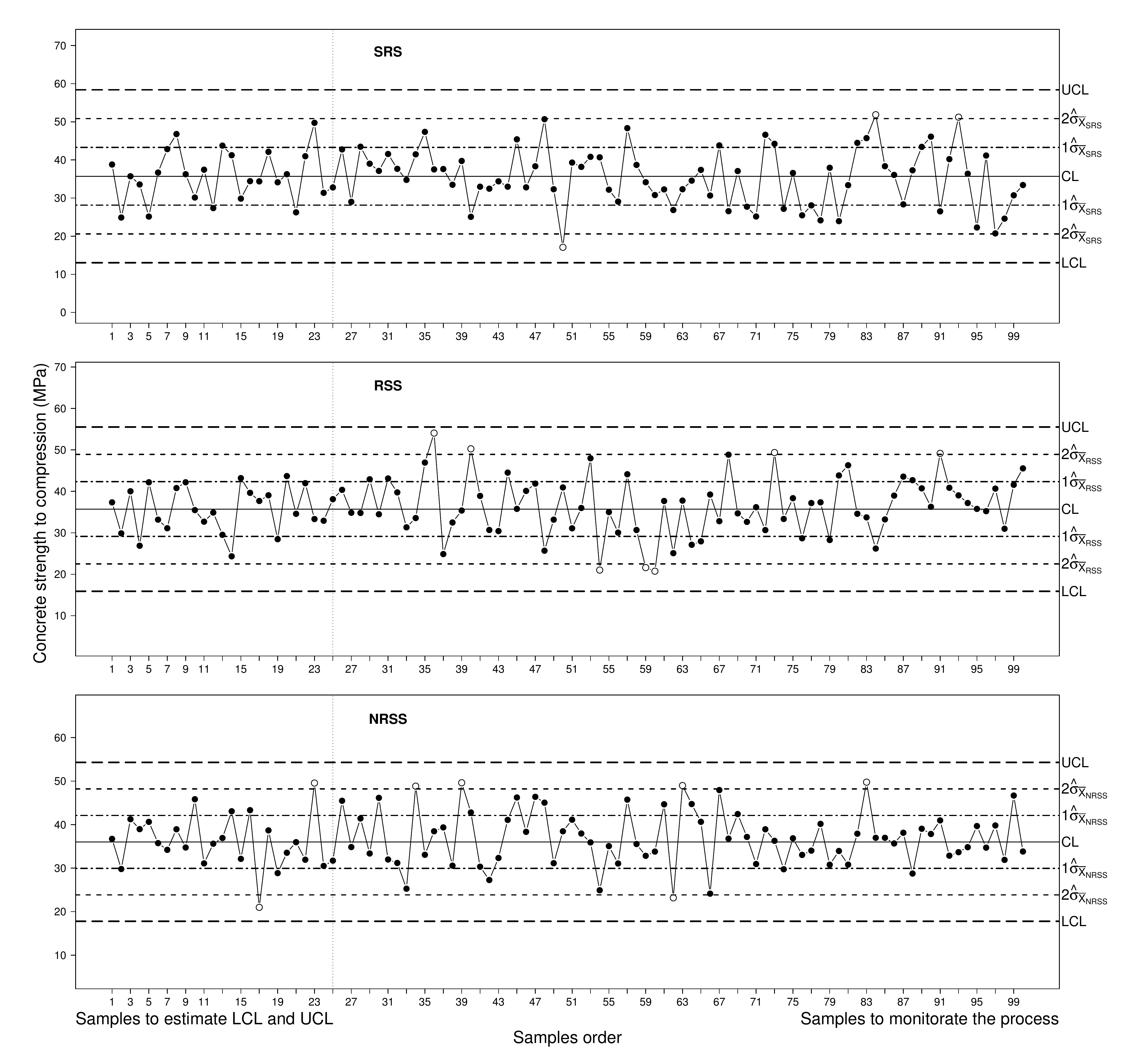}
\caption{Control charts for concrete strength considering $k=5$ and an under control process ($\delta = 0$)} \label{k5delta0}
\end{figure}

\begin{figure}[ht!]
\centering
\includegraphics[width=1\textwidth]{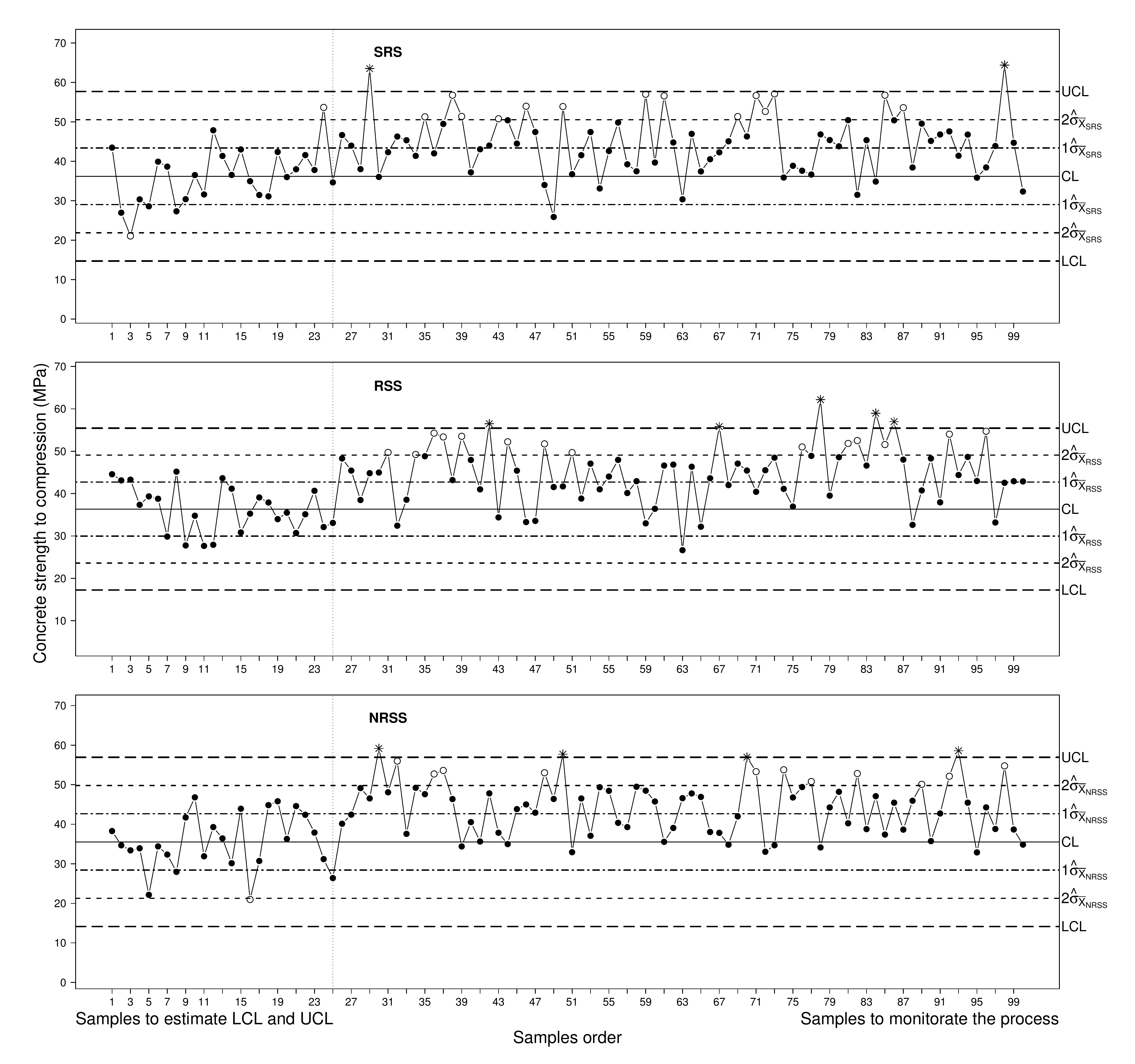}
\caption{Control charts for concrete strength considering $k=5$ and an out of control process ($\delta = 1.2$)} \label{k5delta12}
\end{figure}

Figures \ref{k5delta0} and \ref{k5delta12} present the control charts for $k=5$, under the three sampling designs, considering, respectively, $\delta=0$ and $\delta=1.2$. Once more, it is possible to observe that RSS and NRSS control charts present satisfactory performance, showing randomness and absence of points outside the control limits for an under control process, and also presenting points exceeding the control limits more frequently than SRS in the out of control scenario.


\section{Conclusion}

In this paper, we considered control charts for the mean of a normal distributed process based on NRSS design. These charts were compared to their SRS and RSS based counterparts by means of a simulation study. Under perfect ranking, NRSS control charts overcome all their competitors, providing smaller ARL values for out of control process in all simulated scenarios. In addition, the NRSS control charts showed to be competitive when compared to those based on double ranked set designs. However, such sampling designs require the initial selection of $k^3$ sample units for, after two ordering cycles, selecting a final sample of $k$ units. For example, the ARL for NRSS control charts were smaller in all simulated scenarios when compared to those obtained by EDRSS and DERSS, and surpassed by those provided by DQRSS and QDRSS  when $ k = 5$ \cite{Alo12}. Moreover, this superiority is also verified against DRSS control charts for all considered sample sizes. When considering the DMRSS and MDRSS control charts \citep{Mut03c}, on the other hand, these designs dominate NRSS, providing lower ARL values. However, it should be considered that double ranked set designs could be expensive, and sometimes impracticable, due to a high operational effort. Furthermore, based on its superiority over MRSS, it could be viewed, as a future work, NRSS designs based on two or more ordering cycles.

Under imperfect ranking, we have shown that the efficiency of NRSS control charts becomes smaller as the correlation between the pair of variables decreases. This is a common fact to other designs based on RSS. Even so, the simulated ARL values for NRSS control charts are predominantly smaller (for out of control processes) than for the corresponding ones reached by SRS. Additionally, it was possible to verify the superiority of the NRSS control charts with the ones produced by RSS, MRSS and ERSS in most of the simulated scenarios.

In an illustration with real data regarding concrete strength, the  SRS, RSS and NRSS control charts presented points randomly distributed around the central line, without any points outside the control limits, when we simulated from a process under statistical control. However, for the out of control scenarios, the NRSS and RSS control charts performed expressively better when compared to the usual control charts based on SRS. 

Therefore, based on these results, we recommend the use of NRSS control charts for monitoring the process mean as an efficient alternative to SRS and to other RSS based designs. Under the operational point of view, the ranking of $k^2$ samples units in a single set (instead of ranking $k$ sets of $k$ units, as it occurs in RSS, MRSS and ERSS designs) may, eventually, become a complicating issue, if the ordering criterion is based, for example, on a visual judgment. However, this will usually not make great difference if the ordering criterion is based on an auxiliary variable.

\section*{Disclosure statement}
No potential conflict of interest was reported by the authors.

\newpage



\begin{thebibliography}{99}


\bibitem{Abu04} 
M. Abujiya, and H.N Muttlak, \emph{Quality control chart for the mean using double ranked set sampling}, 
Journal of Applied Statistics 31.10 (2004), pp. 1185--1201.

\bibitem{Alh10} 
S.A. Al-Hadhrami, \emph{Estimation of the population variance using ranked set sampling with auxiliary variable}, 
International Journal of Contemporary Mathematical Sciences. 5.52 (2010), pp. 2567--2576.

\bibitem{Alo11a} 
A.I. Al-Omari, \emph{Estimation of mean based on modified robust extreme ranked set sampling}, 
Journal of Statistical Computation and Simulation 81.8 (2011), pp. 1055--1066.


\bibitem{Alo12}
A.I. Al-Omari, and A. Haq, \emph{Improved quality control charts for monitoring the process mean, using double-ranked set sampling methods}, 
Journal of Applied Statistics 39.4 (2012), pp. 745--763.

\bibitem{Alo14}
A.I. Al-Omari, and C.N. Bouza, \emph{Review of ranked set sampling: modifications and applications}, 
Revista Investigaci\'on Operacional 3 (2014), pp. 215--240.

\bibitem{Als00}
M.F. Al-Saleh, and M.A. Al-Kadiri, \emph{Double-ranked set sampling}, 
Statistics \& Probability Letters 48.2 (2000), pp. 205--212.

\bibitem{Als10}
W.S. Al-Sabah , \emph{Cumulative sum control charts using ranked set sampling data}, Pakistan Journal of Statistics (2010) 26(2).


\bibitem{Bal98}
N. Balakrishnan, and C.R Rao. \emph{Order statistics: theory \& methods}, Elsevier, 1998.

\bibitem{Che03}
Z. Chen, Z. Bai, and B. Sinha, \emph{Ranked set sampling: theory and applications}, 
Vol. 176. Springer Science \& Business Media, 2003.

\bibitem{Che07}
Z. Chen, \emph{Ranked set sampling: its essence and some new applications}, 
Environmental and Ecological Statistics 14.4 (2007), pp. 355--363.

\bibitem{Cos94}
A.F.B. Costa, \emph{(X) over-bar charts with variable sample-size} 
Journal of Quality Technology (1994), pp. 155--163.

\bibitem{Cos07}
A.F.B Costa, and M.S. De Magalhaes, \emph{An adaptive chart for monitoring the process mean and variance}, 
Quality and Reliability Engineering International 23.7 (2007), pp. 821--831.

\bibitem{Del72}
T.R. Dell, and J.L. Clutter, \emph{Ranked set sampling theory with order statistics background}, 
Biometrics (1972), pp. 545--555.

\bibitem{Haq14a}
A. Haq, J. Brown, E. Moltchanova, and A.I. Al-Omari, \emph{Mixed ranked set sampling design}, 
Journal of Applied Statistics 41.10 (2014), pp. 2141--2156.


\bibitem{Haq14c}
A. Haq , \emph{An improved mean deviation exponentially weighted moving average control chart to monitor process dispersion under ranked set sampling}, Journal of Statistical Computation and Simulation (2014) 84(9).

\bibitem{Haq15}
A. Haq, J. Brown, E. Moltchanova and A. I. Al-Omari \emph{Improved exponentially weighted moving average control charts for monitoring process mean and dispersion}, Quality and Reliability Engineering International (2015) 31(2).

\bibitem{Kau95}
A. Kaur, G.P. Patil, A.K. Sinha, and C. Taillie, \emph{Ranked set sampling: an annotated bibliography}, 
Environmental and Ecological Statistics 2.1 (1995), pp. 25--54.

\bibitem{Kho03}
M.B.C. Khoo, \emph{Design of runs rules schemes}, Quality Engineering 16.1 (2003), pp. 27--43.

\bibitem{Kou07}
M.V. Koutras, S. Bersimis, and P.E. Maravelakis. \emph{Statistical process control using Shewhart control charts with supplementary runs rules}, 
Methodology and Computing in Applied Probability 9.2 (2007), pp. 207--224.

\bibitem{Koy17}
N. Koyuncu. \emph{New mean charts for bivariate asymmetric distributions using different ranked set sampling designs}, 
Quality Technology \& Quantitative Management (2017), pp. 1-20.

\bibitem{Mac02}
S.N. MacEachern, Ö. Öztürk, D.A. Wolfe, and G.V. Stark, \emph{A new ranked set sample estimator of variance}, 
Journal of the Royal Statistical Society: Series B (Statistical Methodology). 64.2 (2002), pp. 177--188.

\bibitem{Mci52}
G.A. McIntyre, \emph{A method for unbiased selective sampling, using ranked sets}, 
Crop and Pasture Science. 3.4 (1952), pp. 385--390.

\bibitem{Meh13}
R. Mehmood, M. Riaz, and R.J.M.M. Does. \emph{Control charts for location based on different sampling schemes}, 
Journal of Applied Statistics 40.3 (2013), pp. 483--494.

\bibitem{Mon09}
D.C. Montgomery, \emph{Statistical quality control}, Vol. 7. New York: Wiley, 2009.

\bibitem{Mut97}
H.A. Muttlak, \emph{Median ranked set sampling}, 
Journal of Applied Statistical Sciences 6.4 (1997), pp. 245-255.

\bibitem{Mut03a}
A.H. Muttlak, \emph{Modified ranked set sampling methods}, 
Pakistan Journal of statistics-all series- 19.3 (2003), pp. 315--324.

\bibitem{Mut03b}
H. Muttlak, and W. Al-Sabah, \emph{Statistical quality control based on ranked set sampling}, 
Journal of Applied Statistics 30.9 (2003), pp. 1055--1078.

\bibitem{Mut03c}
H. Muttlak, and A. Mu'azu, \emph{Quality control chart for the mean using double ranked set sampling}, Journal of Applied Statistics 31.10 (2004), pp. 1185--1201.



\bibitem{Sam96}
H.M. Samawi, M.S. Ahmed, and W. Abu-Dayyeh, \emph{Estimating the population mean using extreme ranked set sampling}, 
Biometrical Journal 38.5 (1996), pp. 577--586.

\bibitem{She24}
W.A. Shewhart, \emph{Some applications of statistical methods to the analysis of physical and engineering data} 
Bell System Technical Journal 3.1 (1924), pp. 43--87.


\bibitem{Tak68}
K. Takahasi, and K. Wakimoto, \emph{On unbiased estimates of the population mean based on the sample stratified by means of ordering}, 
Annals of the Institute of Statistical Mathematics. 20.1 (1968), pp. 1--31.


\bibitem{Tag98}
G. Tagaras, \emph{A survey of recent developments in the design of adaptive control charts}, 
Journal of quality technology 30.3 (1998), pp. 212.


\bibitem{WEC58}
Western Electric Company, \emph{Statistical quality control handbook}, The Company, 1958.

\bibitem{yeh98}
I.C. Yeh, \emph{Modeling of strength of high-performance concrete using artificial neural networks},
Cement and Concrete research 28.12 (1998) pp. 1797--1808.

\bibitem{Zam15}
E. Zamanzade, and A.I. Al-Omari, \emph{New ranked set sampling for estimating the
population mean and variance}, Hacettepe Journal of Mathematics and Statistics (2015) 46(92).

\end{thebibliography}
\end{document}